\documentstyle[12pt]{article}

\def\newpic#1{%
   \def\emline##1##2##3##4##5##6{%
      \put(##1,##2){\special{em:point #1##3}}%
      \put(##4,##5){\special{em:point #1##6}}%
      \special{em:line #1##3,#1##6}}}
\newpic{}

\newcommand{\be}{\begin{equation}}
\newcommand{\ee}{\end{equation}}

\newcommand{\hsp}{\mbox{$\hspace{.3in}$}}

\def\bea{\begin{eqnarray}}
\def\eea{\end{eqnarray}}
\def\ba{\begin{array}}
\def\ea{\end{array}}
\def\bac{\begin{array}{c}}
\def\bacc{\left[\begin{array}{cccc}}
\def\baac{\left(\begin{array}{cc}}
\def\eacc{\end{array}\right]}
\def\eaac{\end{array}\right)}

\def\aligned{\bac}
\def\endaligned{\ea}
\def\tag#1{\label{#1}}
\def\text{\rm}
\def\ed{\end{document}}

\def\q#1{\ref{Q#1}}
\def\i#1{\ref{I#1}}

\def\a{\alpha}
\def\b{\beta}
\def\e{\varepsilon}
\def\r{\rho}
\def\rr{r}
\def\g{\gamma}
\def\d{\delta}

\def\k{\psi}

\def\k{\psi}
\def\ps{\k}
\def\ka{\k^{\,\alpha}}
\def\kb{\k^{\,\beta}}
\def\kaa{\k^{\,\a_1}}
\def\kbb{\k^{\,\b_1}}

\def\si{\sigma}
\def\sr{\sigma_{\ttr}}
\def\srr{\sigma_{r}}
\def\sx{\sigma_{r+\hat x}}
\def\sy{\sigma_{r+\hat y}}
\def\sxy{\sigma_{r+\hat x+\hat y}}
\def\sg#1{\sigma^{#1}_{\ttr}}

\def\Q{{\hat Q}}
\def\QQ{{\hat Q}_q}
\def\CR{{\hat {\cal R}}}
\def\V{{\Psi}^{(2)}}
\def\VB{{\Psi}^{(b)}}
\def\CJ{{\hat {\cal J}}}
\def\CI{{\hat {\cal I}}}
\def\CP{{\hat {\cal P}}}
\def\JJ{{ {\cal J}}}
\def\II{{{\cal I}}}

\def\nx{\nabla_{x}}
\def\ny{\nabla_{y}}
\def\mx{\nabla_{-x}}
\def\myy{\nabla_{-y}}
\def\tp{\tilde \psi}

\def\A{{\hat A}}

\def\R{{\hat R}}

\def\VR{\stackrel{*\,\,\,\,}{{ \Psi}^{(2)}}}

\def\P{{ \Psi}^{(b)}}
\def\DP{\stackrel{*}{ \Psi}}

\def\kr{\k^{\,\rho}}

\def\mmy{\nabla_{-y}}
\def\baq{{{\hat {\cal A}}^{(q,s)}}}
\def\bsq{{{\hat {\cal S}}^{\,(q,s)}}}
\def\ba{\hat {\cal A}}
\def\cl{ {\cal L}}
\def\bs{\hat {\cal S}}
\def\K{{\cal K}}
\def\cD{{\cal D}}
\def\caq{{\hat {\cal A}}}
\def\csq{{\hat {\cal S}}}
\def\caa{\,{\cal A}\,}

\def\mp{e^{-ip_1}}
\def\ip{e^{ip_1}}
\def\iq{e^{ip_2}}
\def\mq{e^{-ip_2}}
\def\qq{(p_1\,,p_2)}
\def\mm{(-p_1\,,-p_2)}

\def\ca{{{\hat {\cal A}}_{\,q}}}
\def\cs{{{\hat {\cal S}}_{\,q}}}
\def\s{{SL_{\,q}(2,C)}}
\def\sq{{SL_{\,q}(2)}}
\def\1{\hat {\bf 1}}
\def\0{_{R_{1}}}
\def\2{_{R_{2}}}
\def\J{{\hat J}}

\def\kg{\k^{\,\gamma} }

\def\I{{\hat I}}
\def\kt{{\tilde {\k}}}
\def\ty{\otimes}
\def\tt{\hat \otimes}
\def\bd{\bar {d}}
\def\my{{\dot  \otimes}}

\def\D{{\hat D}}

\def\pa#1{\psi^{\a}_{#1}}
\def\pb#1{\psi^{\b}_{#1}}
\def\pg#1{\psi^{\g}_{#1}}
\def\pro#1{\psi^{\r}_{#1}}

\def\bdp#1{\bar {d} \psi_{#1}}
\def\bpa#1{\bar {d} \psi_{\a}^{#1}}

\def\pai#1{\psi^{\a}_{i}(#1)}
\def\pbj#1{\psi^{\b}_{j}(#1)}
\def\pgk#1{\psi^{\g}_{k}(#1)}
\def\prl#1{\psi^{\r}_{l}(#1)}
\def\dpai#1{\bar {d}\psi^{\a}_{i}(#1)}
\def\dpbj#1{\bar {d}\psi^{\b}_{j}(#1)}
\def\dpgk#1{\bar {d}\psi^{\g}_{k}(#1)}
\def\dprl#1{\bar {d}\psi^{\r}_{l}(#1)}

\def\p#1{\psi^{#1}_{r}}
\def\pp{\psi^{1}_{r^{\prime}}}
\def\ppp{\psi^{2}_{r^{\prime}}}

\def\rf#1{\varphi^{#1}_{r^{\,\prime}}}
\def\f#1{\varphi^{#1}_{r}}
\def\fy#1{\varphi^{#1}_{r+\hat y}}
\def\fx#1{\varphi^{#1}_{r+\hat x}}
\def\fxy#1{\varphi^{#1}_{r+\hat x+\hat y}}
\def\fr{\varphi_{r}}
\def\vr{\varphi}

\def\rp{r^{\prime}}
\def\tr{\tilde r}
\def\trp{{\tilde r}^{\prime}}
\def\ttr{ R}

\def\dtp#1{\bar {d} {\tilde\psi}_{#1}}

\def\px#1{\psi^{#1}_{r+\hat x}}
\def\py#1{\psi^{#1}_{r+\hat y}}
\def\pxy#1{\psi^{#1}_{r+\hat x+\hat y}}

\def\wrr{\omega_{\ttr}}
\def\ttr{\scriptscriptstyle R}

\begin{document}

\begin{titlepage}
\begin{flushright}
ITP-94-57E \\
January, 1995 \\
Revised version \\
March, 1995 \\
hep-th/9501116
\end{flushright}
\vspace{1cm}

\begin{center}
{\large\bf  $(l,q)$-Deformed Grassmann Field and }
\end{center}
\begin{center}
{\large\bf  the Two-dimensional Ising Model}
\end{center}
\vspace{.5cm}
\begin{center} {Anatolij I.~Bugrij\footnote
{E-mail address:
abugrij\@gluk.apc.org }}
\end{center}
\smallskip
\begin{center} {\it Bogolyubov
 Institute for Theoretical Physics,}
\end{center}
\begin{center} {\it 252142 Kiev-142,
Ukraine}
\end{center}
\medskip
\begin{center} {Vitalij N.~Shadura}
\end{center}
\smallskip
\begin{center} {\it Bogolyubov Institute for
 Theoretical Physics,}
\end{center}
\begin{center} {\it 252142 Kiev-142, Ukraine}
\end{center}
\bigskip

\begin{abstract}
  In this paper we
construct the exact representation of the Ising partition function
in the form of the $ SL_q(2,R)$-invariant functional
integral for the lattice free $(l,q)$-fermion field theory ($l=q=-1$).
It is shown that the $(l,q)$-fermionization
allows one to re-express the partition function of the eight-vertex
model in external field through functional integral with
four-fermion interaction.  To construct these
representations, we define a lattice $(l,q,s)$-deformed Grassmann
bispinor field  and extend the
Berezin integration rules to this field.  At $l=q=-1, s=1$ we obtain
the lattice $(l,q)$-fermion  field which allows us to fermionize the
two-dimensional Ising model.  We show that the
Gaussian integral over $(q,s)$-Grassmann variables is expressed
through the $(q,s)$-deformed  Pfaffian which is equal to square root
of the determinant of some matrix at $q=\pm 1, s=\pm 1$.

\end{abstract}

\end{titlepage}

\setcounter{page}{1}

\setcounter{equation}{0}
\section{Introduction}

During the last decade, a considerable progress in understanding
 the meaning of the infinite-dimensional dynamical symmetries  in
 the exactly solvable two-dimensional lattice spin models and quantum
 field theories has been achieved.
 As it was shown by Belavin, Polyakov and
 Zamolodchikov in [1], conformal invariance of the critical
 fluctuations in the two-dimensional lattice spin systems allows one to
 find a universality classes classification
 of the critical behaviour
  of exactly solvable lattice systems
  and to assign to each class
the corresponding two-dimensional quantum conformal
 field theory. This result is connected with the fact that, at
 critical points, such systems, besides the invariance with respect to the
 global conformal transformations ($SL(2,C)$ group) possess larger
  symmetry, that is, the space of eigenvectors of the transfer matrix
 (Fock space for quantum conformal field theories) has an
 invariance with respect to the infinite-dimensional Virasoro algebra.

 However, as soon as the correlation
 radius of critical fluctuations is finite nearby the critical point,
 the conformal invariance is broken.  Nevertheless, as it was shown
 by Zamolodchikov in [2], exactly solvable two-dimensional quantum
 field theory can still possess infinitely many integrals of motion,
in the vicinity of the fixed point (this point corresponds to a quantum
 conformal field theory).
 This allowed him to suggest that the Fock space could
 be realized as an irreducible representation space for some
 infinite-dimensional algebra.  This idea has been explicitly realized for
 exactly solvable massive quantum field theories. For example, in the case of
 quantum Sine-Gordon model, the Fock space has $U_q(\hat {sl_2})$ symmetry
 [3,4] and the deformation parameter $q$ is connected with the coupling
 constant of the model. Also, as it was shown in [3,4] for the
  $SU(2)$-invariant Thirring model, the creation and annihilation operators
  realize a representation of $U_q(\hat {sl_2})$ at $q=-1$.

In [6] there have been also found infinitely many integrals of motion
nearby the critical point of the eight-vertex model,
corresponding to the two non-interacting Ising sub-lattices. In the remarkable
series of papers by Jimbo, Miwa and their collaborators [7,8], it was shown
that
the eigenvectors of the corner transfer-matrix for the six-vertex model form
the irreducible representation space of $U_q(\hat {sl_2})$ at $-1<q<0$.
Moreover, it was demonstrated in [9,10] that the space of eigenvectors of the
corner transfer-matrix for the RSOS models and, in particular, for the
two-dimensional Ising model [10] can possess this symmetry too.

These results allow one to assume that the partition function of the
 two-dimensional exactly solvable lattice model can be expressed
through the partition function of some $SL_q(2,C)$-invariant lattice
$q$-deformed field theory.

 In this paper, we make the first step examining this assumption and
propose a simple example of the construction of such a type
(lattice field theory) in the case of the two-dimensional Ising model, making
use of the exact representation of the Ising partition function as the
$SL_q(2,R)$-invariant functional
integral in the lattice free $(l,q)$-fermion field theory ($l=q=-1$).
Moreover, it is shown that the method of $q$-femionization allows one to
re-express the partition function of the eight-vertex model in external field
through the functional integral with four-fermion interaction.

For the
construction  of these representations we define a lattice
$(l,q,s)$-deformed Grassmann bispinor field\footnote
 {\noindent Note that for $l=e^{i\pi\theta}$ the lattice $(l,q)$-deformed
 field one can consider  as $q$-deformed non-abelian anyonic field} and
extend the Berezin integration rules to this field.  Here $l$ is a
deformation parameter for the commutation relations of "values" of this
field in two arbitrary lattice sites, $q$ is a deformation parameter for
$q$-Grassmann spinor and $s$ is a deformation parameter for commutation
relations  of two $q$-Grassmann spinors in a $s$-deformed bispinor.  At
$l=q=-1, s=1$ we obtain the lattice $(l,q)$-deformed  field which allows
us to fermionize the two-dimensional Ising model (we call it the lattice
$(l,q)$-fermion field).  Moreover, we show that Gaussian integral over
$(q,s)$-Grassmann variables is expressed through the $(q,s)$-deformed
Pfaffian which at $q=\pm 1, s=\pm 1$ is equal to square root of the
determinant of some matrix.

It is a standard point that, in the scaling limit $T\rightarrow T_c,\,\,
a\rightarrow 0$, the partition function of the two-dimensional Ising model
is considered as the partition function of the free massive Majorana
fermion theory [11]. One can obtain this result by
representing the Ising partition
function as the functional integral in the lattice real
fermion field theory [12-16].

The fermionization method proposed in [13,15]
 proved to be effective for solving to the boundary condition problems and
 allows one to calculate the partition function of the two-dimensional
 Ising model with an arbitrary magnetic field attached to the boundaries [16].

 Now we briefly recall the results obtained in  [13,15].
Partition function of the Ising model on the torus is
\be
Z_{I}=\sum_{\{\sigma\}}\,e^{-\b H}=\sum_{\{
\si\}}\exp \left[\,{\K} \,{\sum_{ \rr }}(\,\sigma_{\rr+\hat x} \,\sx
+\sigma_{ \rr+\hat y}\,\sy\,)\right]\,,\label{I1}
\ee
where $r=(x,y)$ denotes the sites of the square lattice of size
 $n\times m$, $x=1,\ldots,n$,
$y=1,\ldots,m$, ${\K}=\b J$, $\sr=\pm 1$, and $\hat x$,  $\hat y$
are unit vectors along the
horizontal $X$ and vertical  $Y$ axes respectively. This partition
function can be presented as the sum of the functional integrals over
the lattice real fermion field
\be
Z_I= Z^{AA} +Z^{AP}+Z^{PA}-Z^{PP}\,,\label{I2}\ee
\be
Z=
{\mbox{const}}\int{\cD}\psi \exp \,\left[ \,{\frac 1 2}
\sum {\ps_r^{i}}\,D^{ i\,j}_{r,r^{\prime}}\,{\ps^{j}_{r^{\prime}}}
\,\right] \,,\label{I3} \ee
 where
 \be
\hat D= \left[\begin{array} {cccc} 0
&1&1+t\nx & 1 \\
\vspace{2truemm}
-1 &0 &1 &1+t\ny \\
\vspace{2truemm}
-1-t\mx &-1 &0 &1\\
\vspace{2truemm}
-1 & -1-t\myy & -1  &0
\ea \right]
\label{I4} \ee
$$
\nx \ps_r=\psi_{r+\hat x},\,\,\ny \ps_r=\psi_{r+\hat
y},\,\,t=\tanh\K, \,\,
\cD\psi={\prod_{ i,r}}{d\p{i}}\,.
$$
Here, for example, the indices
 $P, A$ in  $Z^{PA}$  denote
periodic and antiperiodic
 boundary conditions for the field $\ps_r$ along the
horizontal $X$ and vertical  $Y$ axes respectively.
The appearance of the four-component Grassmann (bispinor) field
 $\p\a$ is related to the requirement of linearity of the
 quadratic form in (\i 3) with respect to the shift operators
 $\nx\,,\,\ny$.
 Note also that for such a quadratic form the doubling problem for the
lattice fermions, arising in the course of usual formulations
of lattice fermion field theories [17] is absent.

 In the momentum representation, the matrix
$\D$ has the following form
\be
D^{i\,j}\qq=
\left[\begin{array} {cccc}
0  &1&1+t\ip & 1 \\
\vspace{2truemm}
-1 &0 &1 & 1+t\iq \\
\vspace{2truemm}
-1-t\mp & -1 &0 &1\\
\vspace{2truemm}
-1& -1 -t\mq
&-1 &0
\ea\right] \,,\label{I5}\ee
where for the periodic and antiperiodic
 boundary conditions  along the $X$, $Y$ axes, $p_1,p_2$ are integer
 and half-integer respectively.

Using momentum representation, it is
not hard to calculate functional integral (\i 3) and, for example, one
 can obtain for the $Z^{AA} $
 \be
 Z^{AA}={\text {const}}\sqrt{\det  \D}
=c\,\prod_{p_1,p_2}(s+s^{-1}-\cos p_1 -\cos p_2)\,,\label{I6}\ee
where $s=\sinh 2\K$ and $c={\mbox {const}}$, $\sinh 2\K_{c}=1$ at
$T=T_{c}$.

Introducing the notation
\be
\cosh\g(p_2)=s+s^{-1}-\cos p_2 \label{I7} \ee
and calculating the product over
$p_1$, one gets
\be
Z^{AA}=c\,\prod_{p}\,\cosh {\frac
m2}\g(p)=c\,\prod_{p}\,\exp(\frac12 ma \frac{\g(p)}a )
\,(1+\exp(-ma\frac{\g(p)}a ))\,.\label{I8}\ee
Here $a$ is a dimensional
lattice spacing, and we define the dimensional momentum by
substitution $p_2\rightarrow p a$.  In order to study the continuum
limit, one expands (\i 7) into a power series in $p$ around $p=0$.
Restricting ourselves to the quadratic terms in $p$, we obtain
\be
\frac{\g^2(p)} {a^2 }=\frac{2(s-1)^2}{sa^2}+p^2=\e^2(p)\,.
\label{I9} \ee
Considering different limits in (\i 9), one gets $\e(p)$
either for the massless or for the massive Majorana fermions:

 \ $ 1). \,\,T=T_c\,,\hsp s=1\,,\quad a\to 0\,,$\

 $$
 \mu=\lim_{\bac {s=1}\\ a\to 0 \ea}\sqrt { \frac 2 s }
 \frac {s-1}a=0 \quad\text{and}\hsp \e^2(p)=p^2\,, $$

  \ $2).\,\, T\rightarrow T_c\,,\hsp s\to 1\,,\quad
a\to 0\,,$\

 $$
 \mu=\lim_{\scriptscriptstyle\bac {s\to 1}\\ a\to 0 \ea}\sqrt { \frac
 2 s} \frac {s-1}a =\sqrt 2 \hsp\mbox{and}\hsp
 \e^2(p)=\mu^2+p^2\,.
 $$
 The free energy of
 the Ising model in the continuum limit turns into the free energy of
 Majorana fermions at finite temperature. To see this, let us define, as
 $a\rightarrow 0\,,\,n\rightarrow \infty\,,\,m\rightarrow \infty$,
 the system volume as $L=na$  and the inverse  temperature of the
 thermostate in which Majorana fermions are placed as
 $\tilde \b={\tilde T}^{-1}=ma$.  It is not hard to show that in the
 thermodynamic limit $(L\rightarrow \infty)$ the main contribution to
 the free energy comes from $Z^{AA}$.
 Then, for the free energy density $f$
  one obtains from (\i 8)
 $$
 f=-\b^{-1}\,\int\limits_{-\infty}^{\infty}dp\,\ln[1+\exp (-{\tilde
 \b}\e(p))]\,.
 $$
 This expression confirms  ("+" in front of the
 exponential) that the partition function of the two-dimensional Ising
 model on the strip of width
 $\tilde \b=ma$  in the continuum limit  describes the Majorana
 fermion system in the thermostat.

 This paper is organized as follows. In section 2 we give a short
 introduction to $q$-spinors and the quantum matrix group
 ${{SL_{\,q}(2,C)}}$. In section 3 we
 define the lattice
$(l,q,s)$-deformed Grassmann bispinor field.
 In section 4 we define  integral calculus for
 this  field. In section 5 we propose a method to
 represent the partition function of the two-dimensional Ising
 model with the nearest-, next-nearest-neighbour and four-spin
 interactions (eight-vertex model in external field) in the form of
 the functional integral over the lattice $(l,q)$-fermion field $(l=q=-1,
 s=1)$.
 To realize this representation we use the
 $GL_q(2,C)$-covariant generalization of the Berezin integration over
 Grassmann field proposed in section 4 and calculate the
  functional integral for the Ising model with the nearest-neighbour
  interaction.

\setcounter{equation}{0}
\section {q-Spinors and the quantum group $SL_q(2,C)$}

Let us briefly recall the known facts about
$q$-spinors,  their $q$-Grassmann realization and the quantum
 matrix group $SL_q(2,C)$ [18-23].
 By $q$-spinor $v$ we mean the
two-component object
$$ v=\left(\bac v^1 \\ v^2 \ea\right) ,
$$
with
the following commutation relation for the components
 \be
v^2
v^1={\frac 1q} v^1 v^2, \label{Q1}\ee
where $q$ is generally a complex
number.  At $q=1$ it is usual spinor. Following the paper [19], one
can look at $q$-spinors as belonging to the quantum plane $A^{2|0}_q$.

One can consider this quantum plane  as the two-dimensional
representation space for the quantum matrix group
$\s$ ($q$-deformed $SL(2,C)$) [18-23].
Using the $\R$-matrix for this group,
one can  rewrite commutation relation (\q 1) in the covariant form
\be
v^{\a}\,v^{\b}={\frac 1q}\R^{\,
 \a\b}_{q\,\g\r}\,v^{\g}\,v^{\r}, \label{Q2}
\ee
where Greek indices go over $1, 2$ and
$\R_q$ is a symmetric $4\times 4$ matrix (the rows
and columns are numbered as (11), (12), (21), (22))
\be
\R^{\,\a\b}_{q\,\g\r}\,=\R_{q\,\a\b}^{\,\g\r},\hsp
\R_q=
\bacc
q&0& 0& 0 \\ 0& q-q^{-1}& 1&  0 \\ 0&  1&  0&  0 \\ 0&  0&  0 & q
\eacc , \label{Q3}
\ee
This matrix has eigenvalues $q$ and $-q^{-1}$ and satisfies the
quantum Yang-Baxter equation [18-20]
\be
\R_{q\,12}\,\R_{q\,23}\,\R_{q\,12}=\R_{q\,23}\,\R_{q\,13}\,\R_{q\,23}
\,,\label{Q4a}
\ee
and the Hecke relation
\be
\R_q^2-(q-q^{-1})\R_q-\I=(\I+q\R_q)(\I-q^{-1}\R_q)=0, \label{Q4b}
\ee
where
$
\I=I^{\,\a\b}_{\g\r}=\d^{\a}_{\g}\,\d^{\b}_{\r} $ is the unit matrix.
One can write the following representation for the $\R$-matrix
[20-22]
$$
\R_q=q\cs -q^{-1}\ca,
$$
where matrices
\be \ca  ={\frac 1{1+q^{-2}}}(\I-q^{-1}\R_q), \label{Q5a}
\ee
\be
 \cs  ={\frac 1{1+q^2}}(\I+q\R_q)  \label{Q5b}
\ee
are orthonormal projectors (to the subspaces with eigenvalues $-q^{-1}$ and
$q$
 respectively)
 $$
\ca \cdot \cs=\cs \cdot \ca =0,\hsp \ca^2=\ca, \hsp \cs^2=\cs.
$$
These projections are the quantum counterparts of the classical
$(q=1)$ antisymmetrizer and symmetrizer for the tensors with two
indeces.

Using the $q$-antisymmetrizer $\ca$, one can rewrite commutation
relations (\q 2)
$$
{{\caq}^{\,\a\b}}_{q\ \g\r}\,v^{\g}\,v^{\r}=0.
$$
The classical $(q=1)$ Grassmann spinor $z^{\a} $
can be defined through
$$
{{\csq}^{\,\a\b}}_{q\ \g\r}\,z^{\g}\,z^{\r}=0,
$$
or, in components,
$$
(z^1)^2=(z^2)^2=0,\hsp z^1\,{z^2}+{z^2}\,z^1=0.
$$
Analogously, one can define the $q$-Grassmann spinor  $\k$ at $q^2\neq
-1$ \ [21,22]
\be
{\csq}^{\,\a\b}_{q\ \g\r}\,\k^{\g}\,\k^{\r}=0, \qquad
\k^{\a}\,\k^{\b}=-q\,{\R^{\,\a\b}_{q\,\g\r}}\,\k^{\g}\,\k^{\r} ,
\label{Q6}
\ee
or, in components,
\be
(\k^1)^2=(\k^2)^2=0,\hsp \k^2\,\k^1=-q\,\k^1\,\k^2. \label{Q7}
\ee
It implies (see the paper [19]) that $\k$
belongs to the quantum plane $A_q^{0|2}$, which we denote as
$\V$.

In order to show that commutation relation (\q 6) is invariant with respect to
the transformation of $\k$ by a quantum matrix $\A\in\s$\
\be
\kt^{\a}=\A^{\a}_{\b}\,\k^{\b}=
\baac { A}^{\,1}_{\,1}&  {
A}^{\,1}_{\,2} \\ { A}^{\,2}_{\,1} & { A}^{\,2}_{\,2}
\eaac\,
\left(\bac \k^1 \\ \k^2 \ea\right)=
\baac a& b \\ c& d
\eaac\,\left(\bac \k^1 \\ \k^2 \ea\right) \,, \label{Q8}
\ee
we recall some properties of $\s$.

The matrix elements $\A^{\a}_{\b}$  commute with the components
 $\k^{\a}$ and belong to the associative algebra of functions on the
quantum group  $\sq $, which we denote as
 $\caa$ or $Fun_q(SL\,(2))$ [20]. This algebra is a Hopf algebra.
It implies that the following maps are defined in this algebra:

a) comultiplication $\bigtriangleup$
$$
\caa {\stackrel {\bigtriangleup}{\longrightarrow}}\caa\ty\caa\,\,:
\hsp \bigtriangleup(A^{\,\a}_{\,\b})=
A^{\,\a}_{\,\g}\ty A^{\,\g}_{\,\b},
$$
where symbol "$\ty$" denotes the tensor product of the quantum spaces,

b) counit $\e$
$$
\caa {\stackrel {\e}{\longrightarrow} } {\bf C} \,\,:\hsp
\e(A^{\,\a}_{\,\b})=\d^{\,\a}_{\,\b},
$$
where ${\bf C}$ is a complex number,

c) antipode $i$
$$
\caa  {\stackrel { i}{\longrightarrow}} \caa \,\,:\hsp
i(A^{\,\a}_{\,\b})=
(-q)^{\a-\b}{\tilde A}^{\,\b}_{\,\a},
$$
where ${\tilde A}^{\,\b}_{\,\a} $ is the quantum minor of the matrix
element $ A^{\,\a}_{\,\b} $
defined for the quantum matrix $\A\in\s$ in the following way [20]
$$
{\tilde A}^{\,1}_{\,1}={ A}^{\,2}_{\,2},\hsp {\tilde A}^{\,2}_{\,2}=
{A}^{\,1}_{\,1},\hsp {\tilde A}^{\,1}_{\,2}={A}^{\,1}_{\,2},\hsp
{\tilde A}^{\,2}_{\,1}={A}^{\,2}_{\,1},
$$

d) multiplication $m$
$$
\caa \ty \caa {\stackrel  {m}{\longrightarrow}} \caa \,\,:\hsp
m(A^{\,\a}_{\,\b} \ty A^{\,\g}_{\,\r})=A^{\,\a}_{\,b}\,
A^{\,\g}_{\,\r},
 $$

and the matrix elements  $A^{\,\a}_{\,\b}$ obey the commutation relations
\be
{\R^{\,\a\b}_{q\,\g\r}}\,A^{\,\g}_{\,\mu}\, A^{\,\r}_{\,\nu}=
A^{\,\a}_{\,\g}\, A^{\,\b}_{\,\r}\,{\R^{\,\g\r}_{q\,\mu\nu}}.\label{Q9}
\ee
which are equaivalent, due to (\q 8), to
$$
 ab=qba,\, ac=qca,\, bd=qdb,\, cd=qdc,\, bc=cb,\,
ad-da=(q-q^{-1})bc\,.
$$

These maps allows one to define the following operations with quantum the
matrix $\A\in\s$:

a)  $\bigtriangleup(\A)=\A\my\A$  defines the rule of multiplication
of quantum matrices, and matrix elements of  $\A\my\A$ have the form
$A^{\,\a}_{\,\g}\ty A^{\,\g}_{\,\b}$ ( $\my $ denotes the tensor
product along with the usual matrix multiplication),

b) $\e(\A)=\hat {\bf 1} $ defines the unit matrix in $\s$,
$\hat {\bf 1}=\d^{\,\a}_{\,\b}$,

c) $i(\A)=\A^{-1}$ defines the inverse matrix, that is
$i(\A)\cdot\A=\A \cdot i(\A)=\hat {\bf 1}$, where "$\cdot$"
denotes the matrix multiplication,

d)  $m(\A\ty\A)=\A\tt\A$ defines the usual tensor product for
quantum matrices, and matrix elements of $ \A\tt\A$
have the form $A^{\,\a}_{\,\b}\, A^{\,\g}_{\,\r}$.

e)  the quantum determinant of the quantum matrix $\A$\
$$
det_q \A =\sum_{\a,\,\b} (-q)^{\b-\a}
A^{\,\a}_{\,\b}\,{\tilde A}^{\,\b}_{\,\a} =
1\,.
$$
Using (\q 9), one immediately shows that it commutes with all the matrix
elements $A^{\,\a}_{\,\b}$. Linear transformation (\q 8) of the
quantum plane $\V$ can be rewritten through the map $\d$\
$$
\V  {\stackrel {\d}{\longrightarrow}} \A\my\V\,\,:\hsp
\d(\k)=\A\my\k, \hsp  \d(\k^{\a})=A^{\a}_{\b}\ty\k^{\b}.
$$
It means that $\V$ is a comodule of  $\s$.
Then, using the map $\d\tt\d$, the action of quantum matrix $\A$ on the
tensor product $\V\ty\V$ can be defined as the action of the tensor product
$\A\tt\A$\
$$
\V\ty\V  {\stackrel {\d\tt\d}{\longrightarrow}} (\A\my\V)\ty(\A\my\V) \,.
$$
It is convenient to define the matrix $\A\tt\A$ in the
following way
\be
\A\tt\A=A_1\cdot A_2,\hsp A_1=\A\tt\1
=A^{\a}_{\b}\,\d^{\g}_{\r},\hsp A_2=\1\tt\A
=\d^{\a}_{\b}\,A^{\g}_{\r}\,.  \label{Q10}
\ee
Hence, its action on $\V\ty\V$ can be written as
$$
(\A\tt\A)\my(\k\ty\k)=A_1\cdot
 A_2\my(\k\ty \k)  = \hsp\,$$
 \be
 (\A\tt\1)\cdot (\1\tt\A)
 \my(\k\ty\k)=(\A\my\k)\ty(\A\my\k)  =\kt\ty\kt.\label{Q11}
\ee
 After using the matrices $A_1$ and $A_2$, commutation relation (\q 9)
can be represented in the form
\be
\R_q\cdot A_1\cdot  A_2=A_1\cdot  A_2\cdot \R_q\,. \label{Q12}
\ee

Let us  show that quantum matrices in (\q 8), (\q 9) form
the quantum matrix group $\s$ with respect to the multiplication $\my$.
It is easy to show that matrix elements of
 $\A^{\prime\prime}=\A^{\prime}\my \A$ satisfy
 commutation
relation (\q 9), if $ \A^{\prime},\, \A \in \s$.
Indeed, one defines matrices
$$
A_1^{\prime}=\A^{\prime}\tt\1,\qquad  A_2^{\prime}=\1\tt
\A^{\prime},
$$
which satisfy the relations
\be
\R_q\cdot A_1^{\prime}\cdot  A_2^{\prime}=A_1^{\prime}\cdot
A_2^{\prime}\cdot \R_q\,, \hsp
A_1^{\prime\prime}=A_1^{\prime}\my A_1, \hsp
A_2^{\prime\prime}=A_2^{\prime}\my A_2. \label{Q13}
\ee
Using the comultiplication property
$$
(A_1^{\,\prime}\ty A_1)\cdot (A_2^{\,\prime}\ty A_2)=
(A_1^{\,\prime}\cdot  A_2^{\,\prime})\my (A_1 \cdot  A_2)
 $$
 and (\q {12}), (\q {13}), we get
$$
\R_q\cdot A_1^{\prime\prime}\cdot
 A_2^{\prime\prime}=\R_q\cdot (A_1^{\prime}\my A_1)\cdot (
 A_2^{\prime}\my A_2)= \R_q\cdot (A_1^{\prime}\cdot  A_2^{\prime})\my
( A_1\cdot A_2)=$$
$$
(A_1^{\prime}\cdot  A_2^{\prime})\my (A_1\cdot
A_2)\cdot \R_q= (A_1^{\prime}\my A_1)\cdot  ( A_2^{\prime}\my
A_2)\cdot \R_q=A_1^{\prime\prime}\cdot  A_2^{\prime\prime}\cdot \R_q .
$$
Hence, one concludes that the matrix elements of
$\A^{\prime\prime}$ do really satisfy relation (\q 9).
Moreover, the $q$-determinant has the property
$$
det_q\A^{\prime\prime}=det_q(\A^{\prime}\my \A)=det_q(\A^\prime)\ty
det_q(\A)\,.
$$
Then, the matrix elements
$A_{\,\,\b}^{\prime\prime\,\a}\in Fun_q(SL\,(2))$.

Now one can easily show
the invariance of commutation relation (\q 6)
under transformation (\q 8).
To see this, it is convenient to rewrite  (\q 6) in the form
$$
(\I+q\R_q)\cdot  (\k\ty\k)=0.
$$
Then, using
(\q {11}) and (\q {12}), one proves the invariance by means of the
following chain of equalities
$$ (\I+q\R_q)\cdot  (\kt\ty\kt)=
(\I+q\R_q)\cdot  (\A\my \k)\ty(\A\my \k)=$$
$$ (\I+q\R_q)\cdot  A_1\cdot
A_2 \my (\k\ty\k)= A_1\cdot  A_2 \cdot (\I+q\R_q)\my (\k\ty\k)=$$
\be
(\A\tt\A)\my (\I+q\R_q)\cdot (\k\ty\k)=0.
 \label{Q14}
\ee

\setcounter{equation}{0}
\section{ Lattice (l,q,s)-Grassmann bispinor field}

In this section  we define the lattice (l,q,s)-deformed Grassmann
bispinor field which is needed for the $(l,q)$-fermionization of the
two-dimensional Ising model.
For definition of this field we  use  the results of the
paper [24].
Recall that $l$ is a deformation parameter of the commutation
relations of this field at two different lattice sites,
$q$ is a deformation parameter for $q$-Grassmann spinor and $s$ is a
deformation parameter for the commutation relations  of two
$q$-Grassmann spinors in $s$-deformed bispinor.

At first let us define $(q,s)$-Grassmann bispinor $\k=\{\k^\a_i\}$
( $\a=1,2$ and $i=1,2$)
\be
\k=\left(\bac
\k_1 \\ \k_2 \ea\right),\hsp   \k_1=\left(\bac \k^1_1 \\
\k^2_1 \ea\right),\hsp \k_2= \left(\bac \k^1_2 \\ \k^2_2 \ea\right) ,
\tag{3.1}
\ee
 where $\k$ belongs to a quantum vector space $\VB$ and $\k_1 \in
 \V_1$,\  $\k_2 \in \V_2$. Then, from (2.8) it follows that $\k_1$
 and $\k_2$ satisfy
 the commutation relations
  \be
\k_i^{\a}\ty\k_i^{\b}=-q\,{\R^{\,\a\b}_{q\,\g\r}}\,\k_i^{\g}\ty\k_i^{\r}\,.
 \tag{3.2}
 \ee
 We consider $\VB$ as the Hecke sum with respect to the
 matrix $\QQ$ [24]
 \be
\VB=\V_{1}{\oplus}_{\Q} \V_{2} \,. \tag{3.3}
\ee
This matrix defines
commutation relations  of  $\k_1$ and
$\k_2$ in $s$-deformed bispinor $\k$.
Following definition of the  Hecke sum [25,26], let us consider the
$\QQ$-matrix as linear map\hfill\break $Q_q\,:\,\,\V_{1}\ty\V_{2}
\longrightarrow \V_{2}\ty\V_{1}\,:$\
\be
Q_q(\pa {1}\ty \pb {2})\equiv \Q^{\,\a\b}_{q\,\g\r}\,\pg
{1}\ty \pro {2}=-\frac 1s \pa {2}\ty \pb {1}\,,
 \tag{3.4}\ee
$$
Q_q^{-1}(\pa {2}\ty \pb {1})\equiv
{(\QQ^{-1})}^{\,\a\b}_{\,\g\r}\,\pg {2}\ty \pro {1}=-s\, \pa {1}\ty
\pb {2} \,,
$$
where in general case the deformation parameter $s$ is a complex
number.

Following  (3.2) and (3.4), we define  the $\R$-matrix
for $\VB$
as linear map \hfill\break
$ \R\,:\,\,
\VB\ty\VB\longrightarrow \VB\ty\VB$:
\be
\R(\pa {i}\ty \pb {j})\equiv \R^{\,\a\b\,k\,l}_{\,\g\r\,i\,j}\,\pg
{k}\ty\pro {l} = - \II^{\,\a\b\,k\,l}_{s\,\g\r\,i\,j}\,\pg {k}\ty\pro
{l}\,, \tag{3.5}
\ee
where $\CI_s$ is a diagonal matrix
\be
\II^{\,\a\b\,k\,l}_{s\,\g\r\,i\,j}=
\frac 1q \d^\a_\g\,\d^\b_\r\, \d^k_i\, \d^l_j\,\d_{i\,j}-
\frac 1s \d^\a_\g\,\d^\b_\r\, \d^k_i\, \d^l_j\,(\d_{i\,j}-1),
\tag{3.6}
\ee
$$
\II^{\,\a\b\,k\,l}_{s\,\g\r\,i\,i}\,\pg {k}\ty\pro {l}=\frac 1q
\,\pa {i}\ty \pb{i},\hsp
\II^{\,\a\b\,k\,l}_{s\,\g\r\,i\,j}\,\pg {k}\ty\pro {l}=\frac 1s
\,\pa {i}\ty \pb{j},\hsp i\neq j\,,
$$
and require
that the $\R$-matrix satisfies the quantum
Yang-Baxter equation
\be
\R_{\,12}\,\R_{\,23}\,\R_{\,12}=\R_{\,23}\,\R_{\,12}\,\R_{\,23}
\,,\tag{3.7}
\ee
and the Hecke relation
\be
\R^2-(\CJ_s-\CI_s)\R-\J_s=(\J_s+\CJ_s\cdot\R)\cdot(\J_s-\CI_s\cdot\R)=0,
\tag{3.8}
\ee
where
$\J_s=J^{\,\a\b\,k\,l}_{s\,\g\r\,i\,j}=
I^{\,\a\b}_{\g\r}\,I^{\,k\,l}_{i\,j}$ is the unit matrix,
 $\CJ_s$ is a diagonal matrix
\be
\JJ^{\,\a\b\,k\,l}_{s\,\g\r\,i\,j}=
q\, \d^\a_\g\,\d^\b_\r\, \d^k_i\, \d^l_j\,\d_{i\,j}-
s\, \d^\a_\g\,\d^\b_\r\, \d^k_i\, \d^l_j\,(\d_{i\,j}-1),
\tag{3.9}
\ee
$$
\JJ^{\,\a\b\,k\,l}_{s\,\g\r\,i\,i}\,\pg {k}\ty\pro {l}=q
\,\pa {i}\ty \pb{i},\hsp
\JJ^{\,\a\b\,k\,l}_{s\,\g\r\,i\,j}\,\pg {k}\ty\pro {l}=s
\,\pa {i}\ty \pb{j},\,\, i\neq j\,,
$$
and $\J_s=\CI_s\cdot\CJ_s$.

It is not hard to check that the following anzats for $R$-matrix
\be
\R^{\,\a\b\,k\,l}_{\,\g\r\,i\,j}=
\R^{\,\a\b}_{q\,\g\r}\,\d^k_i\, \d^l_j\,\d^{k\,l}+
(s-s^{-1}) I^{\,\a\b}_{\g\r}\,\d^k_i\,\d^l_j\,
\Theta^{k\,l}+
\Q^{\,\a\b}_{q\,\g\r}\,\d^k_j\,\d^l_i\,\Theta^{k\,l}
+{(\Q^{-1}_q)}^{\,\a\b}_{\,\g\r}\,\d^k_j\,\d^l_i\,\Theta^{l\,k}
\,, \tag{3.10}
\ee
where
\be
\Theta^{k\,l}=\left\{\bac 1,\,\mbox {if $k<l$}\\ 0,\,\mbox {if $k\ge
l$}\ea
\right. \,\tag{3.11}
\ee
satisfies the Hecke equation (3.8).

In the product $\VB\ty\VB$ in (3.5) we imply the following
ordering:
\be
\VB\ty\VB=
\V_{1}\ty\V_{1}{\oplus}\V_{1}\ty\V_{2}{\oplus}
\V_{2}\ty\V_{1}{\oplus}\V_{2}\ty\V_{2}
 \tag{3.12}
 \ee
 With this ordering the $\R$-matrix has  the form:
\be
\R=\bacc
\R_q & 0 &0
&0 \\
\vspace{2truemm}
0& (s-s^{-1})\I & \QQ & 0 \\
\vspace{2truemm}
0 & \QQ^{-1} & 0 & 0 \\
\vspace{2truemm}
0 & 0 & 0 & \R_q \eacc . \tag{3.13}
\ee

By analogy with (2.6) and (2.7), one can define from (3.8)
  the quantum $(q,s)$-anti-\=symmetrizer
\be
\baq  =(\J_s-\CI_s\cdot \R), \tag{3.14}
 \ee
 and  the quantum $(q,s)$-symmetrizer
 \be
 \bsq  =(\J_s+\CJ_s\cdot\R)  \tag{3.15}
\ee
which are orthogonal projectors
 $$
\baq \cdot \bsq=\bsq \cdot \baq =0,\hsp
(\baq)^2=\CI_s^2\cdot(\CI_s+\CJ_s)\cdot\baq,
$$
$$
(\bsq)^2=\CJ_s^2\cdot(\CI_s+\CJ_s)\cdot\bsq.
$$
The quantum
symmetrizer $\bsq$ allows one to define the commutation relations for the
   $(q,s)$-Grassmann bispinor $\pa i$ $(q^2\neq -1)$\
\be
\bsq\cdot (\psi\ty\psi)=0,\hsp\mbox{or}\hsp
\pa {i}\ty\pb {j}+
(\CJ_s\cdot\R)^{\,\a\b\,k\,l}_{\,\g\r\,i\,j}\,\pg {k}\ty\pro {l}=0\,.
\tag{3.16}
\ee
Let us require that this commutation relations are covariant with
respect to the transformation
 $\psi$ by a quantum matrix $\A\in\s$.
 Then taking into account  (2.4), it is not hard to find  the
 following solution of the quantum Yang-Baxter equation (3.7)
 \be
 \QQ=\R_q\,. \tag{3.17}
 \ee
 Substituting  this solution
 in (3.16), we get
 \be
\aligned
\pa{i} \ty\pb{i} = -q\, \pb{i} \ty \pa{i}\, , \hsp
(\pa{i})^2=0,\hsp\\
\vspace{2truemm}
\pa{2} \ty\pa{1} = -s\,q \pa{1} \ty \pa{2}\, ,\\
\vspace{2truemm}
\k^2_{2} \ty\k^1_{1} = -s \, \k^1_{1} \ty\k^2_{2} \, ,\\
\vspace{2truemm}
\k^1_{2}
\ty\k^2_{1}  = - s  \, \k^2_{1} \ty\k^1_{2} -s(q-q^{-1}) \k^1_{1}
\ty\k^2_{2}\,.
\endaligned     \tag{3.18}
\ee

To define the lattice $(l,q,s)$-Grassmann bispinor field
$\pai {r}$
we consider the d-dimensio-\-nal hypercubic lattice. The lattice sites
are labeled by the vector $ r=(p_1,\,p_2,\,.\,.\,.\, ,\,p_d )$, where
$p_i$ are  integer numbers and the total number of sites is  $N$.
  For the sake of convenience, the lattice spacing $a$ is fixed to be
equal to unity.  Let $\k^{\a}_i (r^{\,\prime})\in \VB_{r^{\,\prime}}$
be the $(q,s)$-Grassmann bispinor on the site $r^{\,\prime}$, which
we consider as  a "value" of the $(l,q,s)$-Grassmann  field $\pai {r}$
on the site $r^{\,\prime}$.  The definition of
the lattice $(l,q,s)$-Grassmann bispinor field requires some
 information about its commutation relation in two arbitrary sites
 $r_m=(l_1,\,l_2,\,.\,.\,.\, ,\,l_d )$ and $r_n=(k_1,\,k_2,\,.\,.\,.\,
,\,k_d )$.  In
this paper we suppose the quasi-one-dimensional ordering for the
lattice sites:  $r_m>r_n$, if $l_1-k_1>0$ and the others $l_i-k_i$ are
arbitrary; if $l_1-k_1=0$, then $r_m>r_n$, if $l_2-k_2>0$  and so on,
and  $ r_1\,<\,r_2\,<\,r_3\,<\,...<\,r_N
 \,\,(m,n=1,...,N)$.

Let us denote a quantum vector space which is generated
by $4N$-component $(l,q,s)$-Grassmann vector $\pai r$
as $\Psi$ ($N$ is the  number of sites on the lattice).
By analogy with (3.3), we must  consider
the quantum vector space $\Psi$ as the Hecke sum  with respect to  some
matrix $\Q$
\be
\Psi=\VB_{r_1}{\oplus}_\Q
\VB_{r_2}{\oplus}_\Q\cdot\cdot\cdot{\oplus}_\Q \VB_{r_{N-1}}{\oplus}_\Q
\VB_{r_N}\,. \tag {3.19}
\ee

For definition of   this sum  and its
$\CR$-matrix let us consider  $\Q$-matrix  as
 linear map $\Q\,:\,\,\P_{r_n}\ty\P_{r_m}
\longrightarrow \P_{r_m}\ty\P_{r_n}$ \
\be
\Q(\pai {r_n}\ty \pbj {r_m})\equiv
\Q^{\,\a\b\,k\,l}_{\,\g\r\,i\,j}\,\pgk {r_n}\ty \prl {r_m}=-\frac 1l
\pai {r_m}\ty \pbj {r_n}\,,\tag{3.20}
\ee
$$
{(\Q^{-1})}^{\,\a\b\,k\,l}_{\,\g\r\,i\,j}\,\pgk {r_m}\ty \prl
{r_n}=-l\, \pai {r_n}\ty \pbj {r_m} \,.
$$
where
$r_m>r_n\,\,(m>n) $ .

By analogy with (3.6)
 we can define the $\CR$-matrix as linear map $ \CR\,:\,\,
\Psi\ty\Psi\longrightarrow \Psi\ty\Psi$:
\be
\CR(\pai {r}\ty \pbj
{\rp})\equiv
\CR^{\,\a\b\,k\,l}_{\,\g\r\,i\,j}(r,\rp |\tr ,\trp)
\,\pgk {\tr}\ty\prl {\trp} =
- \II^{\,\a\b\,k\,l}_{\,\g\r\,i\,j}(r,\rp |\tr ,\trp)
\,\pgk {\tr}\ty\prl {\trp},
\tag{3.21}
\ee
where we imply  summation over values $r_1,\,r_2,\,r_3,\,...,r_N $
of the discrete variables $r,\rp,\tr,\trp$ and
$\CI$ is the diagonal matrix
\be
\II^{\,\a\b\,k\,l}_{\,\g\r\,i\,j}(r,\rp |\tr ,\trp)=
\II^{\,\a\b\,k\,l}_{s\,\g\r\,i\,j} \,\d(r,\tr)\, \d(\rp,\trp)
\,\d(r,\rp) - \frac 1l\,
J^{\,\a\b\,k\,l}_{s\,\g\r\,i\,j}\,\d(r,\tr)\, \d(\rp,\trp) \,(\d(r,\rp)-1)\,.
  \tag{3.22}
\ee

Let us require  that
 the $\CR$-matrix satisfies the quantum
Yang-Baxter equation
\be
\CR_{\,12}\,\CR_{\,23}\,\CR_{\,12}=\CR_{\,23}\,\CR_{\,12}\,\CR_{\,23}
\,,\tag{3.23}
\ee
and the Hecke relation
\be
\CR^2-(\CJ-\CI)\CR-\J=(\J+\CJ\cdot\CR)\cdot(\J-\CI\cdot\CR)=0,
\tag{3.24}
\ee
where
$\J=J^{\,\a\b\,k\,l}_{\,\g\r\,i\,j}(r,\rp |\tr ,\trp)=
J^{\,\a\b\,k\,l}_{s\,\g\r\,i\,j}
\d(r,\tr)\, \d(\rp,\trp)$ is unit matrix,
 $\CJ$ is  diagonal matrix
\be
\JJ^{\,\a\b\,k\,l}_{\,\g\r\,i\,j}(r,\rp |\tr ,\trp)=
\JJ^{\,\a\b\,k\,l}_{s\,\g\r\,i\,j} \,\d(r,\tr)\, \d(\rp,\trp)
\,\d(r,\rp) - l\,
J^{\,\a\b\,k\,l}_{s\,\g\r\,i\,j}
\,\d(r,\tr)\, \d(\rp,\trp) \,(\d(r,\rp)-1)
 \tag{3.25}
\ee
and $\J=\CI\cdot\CJ$.

The $\CR$-matrix satisfying the Hecke equation (3.24)
has the following block
structure:
$$  \CR^{\,\a\b\,k\,l}_{\,\g\r\,i\,j} (r,\rp |\tr ,\trp)=
\R^{\,\a\b\,k\,l}_{\,\g\r\,i\,j}\,\d(r,\tr)\, \d(\rp,\trp)\,\d(r,\rp)+
(l-l^{-1}) J^{\,\a\b\,k\,l}_{\g\r\,i\,j}\,\d(r,\tr)\, \d(\rp,\trp)
\Theta^{r\,\rp}+
$$
\be
\Q^{\,\a\b\,k\,l}_{\,\g\r\,i\,j}\,\d(r,\trp)\,
\d(\rp,\tr)\,\Theta^{r\,\rp}
+\Q^{\,\a\b\,k\,l}_{\,\g\r\,i\,j}\,\d(r,\trp)\,
\d(\rp,\tr)\,\Theta^{\rp\,r}
\,,
\tag{3.26}
\ee
where the $\R$-marix,  and $\Theta^{r\,\rp} $
were defined in (3.10)  and (3.11) respectively,
$\d(r,\tr)$  is  a discrete
$\d$-function.

In the product $\Psi\ty\Psi$ in (3.21) we imply the
following ordering:
$$
\Psi\ty\Psi=
\P_{r_1}\ty\P_{r_1}{\oplus}\P_{r_1}\ty\P_{r_2}{\oplus}
\P_{r_2}\ty\P_{r_1}{\oplus}\P_{r_2}\ty\P_{r_2}{\oplus}
\P_{r_1}\ty\P_{r_3}\hsp
$$
\be
{\oplus}\P_{r_3}\ty\P_{r_1}{\oplus}
\P_{r_2}\ty\P_{r_3}{\oplus}\P_{r_3}\ty\P_{r_2}{\oplus}
\P_{r_3}\ty\P_{r_3}{\oplus}\cdot\cdot\cdot{\oplus}
 \P_{r_N}\ty\P_{r_N}.  \tag{3.27}
 \ee
With this ordering the $\CR$-matrix has block structure and, for
example, for two arbitrary sites $r_n$ and $r_m\,\,(r_n<r_m)$
the block $\tilde {\cal R}_{(n\,m)}$ of the $\CR$-matrix acting
on the sum
$\P_{r_n}\ty\P_{r_n}{\oplus}\P_{r_n}\ty\P_{r_m}{\oplus}
\P_{r_m}\ty\P_{r_n}{\oplus}\P_{r_m}\ty\P_{r_m}$
has the form:
\be
\tilde {\cal R}_{(n\,m)}=\bacc
\R & 0 &0
&0 \\
\vspace{2truemm}
0& (l-l^{-1})\J_s & \Q & 0 \\
\vspace{2truemm}
0 & \Q^{-1} & 0 & 0 \\
\vspace{2truemm}
0 & 0 & 0 & \R \eacc . \tag{3.28}
\ee

By analogy with (3.14) and (3.15) from (3.24) we can define
  the quantum $(l,q,s)$-antisymmetrizer
\be
\ba  =(\J-\CI\cdot \CR), \tag{3.29}
\ee
 and  the quantum $(l,q,s)$-symmetrizer
\be
 \bs  =(\J+\CJ\cdot\CR)  \tag{3.30}
\ee
which are orthogonal projectors
\be
\ba \cdot \bs=\bs \cdot \ba =0,\hsp(\ba)^2=\CI^2\cdot(\CI+\CJ)\cdot\ba ,
\hsp(\bs)^2=\CJ^2\cdot(\CI+\CJ)\cdot\bs . \tag {3.31}
\ee
The
quantum symmetrizer $\bs$ permits to define the commutation relations for
the lattice  $(l,q,s)$-Grassmann bispinor field $\pai r$ $(q^2\neq -1)$\
\be
\bs\cdot (\psi\ty\psi)=0,\,\,\,\mbox{or}\,\, \,\,
\pai {r}\ty\pbj {\rp}+
(\CJ\cdot\CR)^{\,\a\b\,k\,l}_{\,\g\r\,i\,j}(r,\rp |\tr ,\trp)
\,\pgk {\tr}\ty\prl {\trp}=0\,.
\tag{3.32}
\ee
Let us require that
these commutation relations are covariant
with respect to the transformation  of lattice field $\pai r$  by
  a quantum matrix $\A\in\s$  and
 $\psi (r)$ is spinor with respect to  $SL_s(2,C)$.
These requirements lead to  the following solution of the
quantum Yang-Baxter equation  (3.23)
\be
\Q=\Q_q\tt\Q_s\hsp{\text or}\hsp
\Q^{\,\a\b\,k\,l}_{\,\g\r\,i\,j} =
     \Q^{\,\a\b}_{q\,\g\r}\,\Q^{\,k\,l}_{s\,i\,j}\,, \tag{3.33}
 \ee
     where
 $\Q_q$ is defined in (3.17) and
 $$
 \Q_s=\R_s=
 \bacc s & 0 &0
&0 \\ \vspace{2truemm} 0&s-s^{-1}  & 1 & 0 \\ \vspace{2truemm} 0 & 1 &
0 & 0 \\ \vspace{2truemm} 0 & 0 & 0 & s \eacc .
$$

Substituting this solution in (3.32),
we obtain commutation relations
$$
\pai {r_m}\ty\pbj {r_m}=-
(\CJ_s\cdot\R)^{\,\a\b\,k\,l}_{\,\g\r\,i\,j}\,\pgk {r_m}\ty\prl
{r_m}\,,
$$\be
\aligned
\pai{r_m} \ty\pai{r_n} = -l\,q\,s\, \pai{r_n} \ty \pai{r_m}\,,\\
\vspace{2truemm}
\pa i {(r_m)} \ty\pb i {(r_n)} = -l\,s\, \pb i {(r_n)}\ty\pa i
{(r_m)}\,,\\
\vspace{2truemm}
\pa i {(r_m)} \ty\pa j {(r_n)} = -l\,q\, \pa j
{(r_n)}\ty\pa i {(r_m)}\,,\\
\vspace{2truemm}
\pa i {(r_m)} \ty\pb j {(r_n)} = -l\, \pb j {(r_n)}\ty\pa i
{(r_m)}\,,\\
\vspace{2truemm}
\pb i {(r_m)} \ty\pa i {(r_n)} = -l\,s\,\pa i {(r_n)}\ty\pb i{(r_m)}
-l\,s(q-q^{-1})\pa i {(r_n)}\ty\pb i{(r_m)}\,,\\
\vspace{2truemm}
\pa j {(r_m)} \ty\pa i {(r_n)} = -l\,q\,\pa i{(r_n)}\ty\pa j {(r_m)}
-l\,q(s-s^{-1})\pa j{(r_n)}\ty\pa i {(r_m)}\,,\\
\vspace{2truemm}
\pa j {(r_m)} \ty\pb i {(r_n)} = -l\,\pb i {(r_n)}\ty\pa j{(r_m)}
-l(s-s^{-1})\pb j {(r_n)}\ty\pa i{(r_m)}\,,\\
\vspace{2truemm}
\pb i {(r_m)} \ty\pa j {(r_n)} = -l \,\pa j {(r_n)}\ty\pb i{(r_m)}
-l(q-q^{-1})\pb j {(r_n)}\ty\pa i{(r_m)}\,,
 \endaligned
 \tag{3.34}
\ee
where $r_n<r_m$ , $i>j$ and $\a>\b$.

\setcounter{equation}{0}
\section{ Integral calculus
for the lattice $(l,q,s)$-Grassmann field}

At first let us  briefly recall the definiton of
the $GL_q(2,C)$-covariant generalization of the Berezin integration
over the quantum vector space $\V$ [24].

 Let $\VR$ be space of the linear functional on $\P$
 with dual nilpotent basis $\{\bdp{\a }\}$ $((\bdp{\a})^2=0)$
which maps $\V$ in
complex numbers:
 $$
 (\VR,\V)  \longrightarrow {\bf C}
 $$
 and
 \be
 (\bdp{\a}, \kb) =\int\bdp{\a } \kb  =
 \d^\b_\a \, \hsp (\bdp{\a }, c  )=\int\bdp{\a }\, c = 0,
 \tag{4.1}
\ee
 where $c$ is a complex number.

Let us define the action of tensor product of the linear functionals
by the following  relation:
\be
(\bdp{\mu }\ty \bdp{\nu },\,\kg  \ty\kr  )=
(\bdp{\mu },\kg) \ty (\bdp{\nu },\kr  )=
  \int\bdp{\mu }\ty
\bdp{\nu }\,\kg  \ty\kr   =   \d^\g_\mu
\,\d^\r_\nu\,(\d_{\mu\,\nu}-1).  \tag {4.2}
\ee
Using this definition and the following ansats for commutation
relations for $\bdp {\a }$:
$$
\bdp{\a_1 }\ty \bdp{\b_1 }=x\,\bdp{\b_1 }\ty \bdp{\a_1 },
$$
where
$\a_1>\b_1$,
 we obtain (no summing)
$$
\int \bdp{\a_1 }\ty \bdp{\b_1 } \kaa \ty  \kbb  =1=-q
\int \bdp{\a_1 }\ty \bdp{\b_1 } \kbb  \ty  \kaa  =
$$
$$
 -qx
\int \bdp{\b_1 }\ty \bdp{\a_1 } \kbb  \ty  \kaa  = -qx.
$$
{}From here $ x=-\frac 1q$ and
$$
\bdp{\a_1 }\ty \bdp{\b_1 }=-\frac 1q\,\bdp{\b_1 }\ty \bdp{\a_1 },
$$
or in covariant form
\be
\bdp{\a }\ty\bdp{\b }+
q\,(\R^{\,\prime}_q)_{\,\a\b}^{\,\g\r}\,\bdp {\g
}\ty\bdp{\r}=0\,,\tag{4.3}
\ee
where
$(\R^{\,\prime}_q)_{\,\a\b}^{\,\g\r}={\R}_{q\,\b\a}^{\,\r\g}$
 determines commutation relations for matrix elements of inverse
matrix $\A^{-1}\in GL_q(2,C)$\,
\be
{(\R^{\,\prime}_q)^{\,\a\b}_{\,\g\r}}\,(A^{-1})^{\,\g}_{\,\mu}\,
(A^{-1})^{\,\r}_{\,\nu}= (A^{-1})^{\,\a}_{\,\g}\,
(A^{-1})^{\,\b}_{\,\r}\,{(\R^{\,\prime}_q)^{\,\g\r}_{\,\mu\nu}}\,.\tag{4.4}
\ee
Hence, it follows:
\be
\mbox{if}\hsp{\tilde \psi}^\a_i= A^\a_\b \ty \kb \,,\hsp
\mbox{then}\hsp
\dtp {\a }= (A^{-1})^\a_\b \ty \bdp {\b }.
\tag {4.5}
\ee

Using (4.1) and (4.2), one can  define the $GL_q(2,C)$-invariant
Berezin integral on $\V$:
\be
\int \bdp{1 }\ty \bdp{2}
\,\k^1\ty\k^2=1\,.\tag{4.6}
\ee
It is not hard to show that this integral is invariant with respect to
transformations (4.5):
\be
\int \dtp{1 }\ty \dtp{2 }
\, \tp^1\ty\tp^2
=det_q(\A^{-1})\,det_q{\A} \int \bdp{1 }\ty
\bdp{2 }
\,\k^1\ty\k^2=1\,.
 \tag{4.7}
\ee
Using (4.2) and (4.6), one can calculate the integral
\be
\int \bdp{1 }\, \bdp{2 }
\, \ka\, \kb
=\e^{\,\a\b} \,,\tag{4.8}
\ee
where
 \be
 \e^{\,\a\b}=\baac 0 & 1 \\  -q& 0 \eaac\,. \tag{4.9}
 \ee
and
for brevity  the notation of tensor product is omitted.

Now let us consider the integral calculus
for the lattice
$(l,q,s)$-Grassmann
 bispinor field $\k=\{\k^{\a}_i(r)\}$ which satisfy
commutation relations  (\ref{3.34}).
  By
analogy with (4.1) let us define
a space $\DP$ of the linear functional on $\Psi$ (\ref{3.19})
with dual nilpotent basis  $\{\bdp{\a }^i (r)\}$ $([\bdp{\a }^i
 (r)]^2=0)$:
 $$ (\DP,\Psi)  \longrightarrow {\bf C}
 $$ and
 \be
 (\bpa{i}(r), \pb j (r^{\prime}))=\int\bpa{i }(r)\,\pb j (r^{\prime})=
 \d^\b_\a\,\d^i_j\,\d(r,r^{\prime})\,\hsp (\bpa{i}(r), c  )=\int\bpa{i}(r)
 \, c = 0, \tag{4.10}
  \ee
   where $c$ is a complex number.

 Elements of the dual basis satisfy commutation relations
\be
\dpai {r}\ty\dpbj {\rp}+
(\CJ\cdot\CR^{\prime})^{\,\a\b\,k\,l}_{\,\g\r\,i\,j}(r,\rp |\tr ,\trp)
\,\dpgk {\tr}\ty\dprl {\trp}=0\,,\tag{4.11}
 \ee
where $\CR^{\prime}$-matrix has blok structure (\ref{3.28}) and contains
matrices $(\R^{\,\prime}_q)_{\,\a\b}^{\,\g\r}={\R}_{q\,\b\a}^{\,\r\g}$
and
\hfill\break$(\R^{\,\prime}_s)_{\,i\,j}^{\,k\,l}={\R}_{s\,j\,i}^{\,l\,k}$.

Let us define
 the $GL_q(2,C)$-covariant integration measure $\cD\k$
for Berezin integral over the quantum vector space $\Psi$:
\begin{eqnarray}
\cD\k& =
&\bd\k^{1}_1{(r_1)}\,\bd\k^1_2{(r_1)}\,\bd\k^2_1{(r_1)}\,\bd\k^2_2{(r_1)}
\bd\k^{1}_1{(r_2)}\,\bd\k^1_2{(r_2)}\,\bd\k^2_1{(r_2)}\,\bd\k^2_2{(r_2)}
\cdot\cdot\cdot\nonumber\\
\vspace{4truemm}
&\,&\bd\k^{1}_1{(r_N)}\,\bd\k^1_2{(r_N)}\,\bd\k^2_1{(r_N)}\,\bd\k^2_2{(r_N)}
\,.\tag{4.12}\end{eqnarray}
By analogy with (4.8) and (4.9) we can define  $GL_q(2,C)$-invariant
generalization of the Berezin integration for the  lattice
 $(l,q,s)$-Grassmann field
\begin{eqnarray}
\int\cD\k&
\k^{\a_1}_{i_1}(r_1)\, \k^{\a_2}_{i_2}(r_1)\,
\k^{\a_3}_{i_3}(r_1)\, \k^{\a_4}_{i_4}(r_1)\,
...\,
\k^{\d_1}_{j_1}(r_N)\, \k^{\d_2}_{j_2}(r_N)\,
\k^{\d_3}_{j_3}(r_N)\, \k^{\d_4}_{j_4}(r_N)=\nonumber\\
&\e^{\,\a_1\a_2 \a_3\a_4 ...\,\d_1\d_2 \d_3
\d_4 }_{\,i_1\,i_2 \,i_3\,i_4 ...\,j_1\,j_2 \,j_3\,j_4}
(r_1,\,...,\,r_N)\,.\tag{4.13} \end{eqnarray}
Here the tensor $\hat \e$ is defined by the following rules:
$$
\e^{\,1\,2\, 1\,2 ...\,1\,2 \,1\,2 }_{\,1\,1\, 2\,2 ...\,1\,1 \,2\,2 }
(\,r_1,\, ...,\,r_N)=1\,,
$$
and the remaining components are defined by the coefficients appearing
in the l.h.s of the relation
$$
\k^{\a_1}_{i_1}(r_1)\, \k^{\a_2}_{i_2}(r_1)\, \k^{\a_3}_{i_3}(r_1)\,
\k^{\a_4}_{i_4}(r_1)\, ...\,
\k^{\d_1}_{j_1}(r_N)\, \k^{\d_2}_{j_2}(r_N)\, \k^{\d_3}_{j_3}(r_N)\,
\k^{\d_4}_{j_4}(r_N)=
$$
$$
\e^{\,\a_1\a_2 \a_3\a_4 ...\,\d_1\d_2 \d_3
\d_4 }_{\,i_1\,i_2 \,i_3\,i_4 ...\,j_1\,j_2 \,j_3\,j_4}
(r_1,\,...,\,r_N)\,
\k^{1}_{1}(r_1)\,
\k^{2}_{1}(r_1)\, \k^{1}_{2}(r_1)\, \k^{2}_{2}(r_1)\,
...\, \k^{1}_{1}(r_N)\, \k^{2}_{1}(r_N)\,
\k^{1}_{2}(r_N)\, \k^{2}_{2}(r_N)
$$
after reordering the l.h.s. to
the r.h.s. using commutation relations (3.34) for each
specific set of values of the indices.

Now let us  consider  the one-site Gaussian integral over lattice
$(l,q,s)$-Grassmann field.
In order to calculate this integral,
  one uses the following definition
of  $(q,s)$-Pfaffian (which is an extension of the definition of
 usual Pfaffian proposed in [27] to $q$-deformed
commutation relations).

   Consider the quadratic form
$$
w=\,
a^{1\,2}_{1\,1}\,\ps^1_1\,\ps^2_1+a^{1\,1}_{1\,2}\,\ps^1_1\,\ps^1_2+
a^{1\,2}_{1\,2}\,\ps^1_1\,\ps^2_2+a^{2\,1}_{1\,2}\,\ps^2_1\,\ps^1_2+
$$
$$
 a^{2\,2}_{1\,2}\,\ps^2_1\,\ps^1_2+a^{1\,2}_{2\,2}\,\ps^1_2\,\ps^2_2=
{\frac
12}\sum_{\a,\,\b,\,i,\,k}\,a^{\a\b}_{i\,k}\,\ps^\a_i\,\ps^\b_k\,,
$$
where the matrix elements $a^{\a\b}_{i\,k}$ are chosen to
be commutative and the matrix $\hat a$ has the form
$$
\hat a =
\bacc
0 & a^{1\,2}_{1\,1} &a^{1\,1}_{1\,2} & a^{1\,2}_{1\,2}\\
\vspace{3truemm}
-\frac 1q a^{1\,2}_{1\,1}& 0 &a^{2\,1}_{1\,2} &a^{2\,2}_{1\,2}\\
\vspace{3truemm}
-\frac 1{sq} \,a^{1\,1}_{1\,2} & -\frac 1s \,a^{2\,1}_{1\,2 } & 0 &
a^{1\,2}_{2\,2}\\
\vspace{3truemm}
- \frac 1{s}  a^{1\,2}_{1\,2}-(q-q^{-1})a^{2\,1}_{1\,2 }& -\frac 1{sq}
\,a^{2\,2}_{1\,2}& -\frac 1q \,a^{1\,2}_{2\,2}&0 \eacc\, .
$$
Here the lower-triangle matrix elements are determined using
relations (\ref{3.18}).  Let us define $(q,s)$-Pfaffian of $\hat a$
through
$$
\frac 12 w^2= Pf_{(q,s)} (\hat a)\,\ps^1_1 \,\ps^2_1
\,\ps^1_2\,\ps^2_2\, .  $$
Then, taking into account commutation
relations (\ref{3.18}), we get
\be
Pf_{(q,s)}(\hat a)={\frac 12}(1+q^2 s^4)a^{1\,2}_{1\,1}\,a^{1\,2}_{2\,2}-
{\frac s{2}}(1+q^2)a^{1\,1}_{1\,2}\,a^{2\,2}_{1\,2}+s\,q^2
a^{1\,2}_{1\,2}\, a^{2\,1}_{1\,2}\,.   \tag{4.14}
\ee
Using (4.13) for the one site integral, it is immediate to show   that
\be
\int \bd\ps^1_1 \, \bd\ps^1_2 \, \bd\ps^2_1 \, \bd\ps^2_2\,\exp
\left\{{\frac 12}
\sum_{\a,\,\b,\,i,\,k}\,a^{\a\b}_{i\,k}\,\ps^\a_i\,\ps^\b_k\right \} =
Pf_{(q,s)}(\hat a).\tag{4.15}
\ee
Now let us show that, at $q=\pm 1, s=\pm 1$,
\be
Pf_{(q,s)}(\hat a) = \sqrt{ det (\hat b)}\,,\tag{4.16}
\ee
where $det$ is the usual ($q=1$) determinant.
To determine the matrix $ \hat b$ we consider $(q,s)$-bispinor $v$,
which satisfies commutation relations
$$
v^{\a}_i\,v^{\b}_j={\frac 1q}
\R^{\,\a\b\,k\,l}_{\,\g\r\,i\,j}\, v^{\g}_k\,v^{\r}_l
\,,
$$
and require
$$
  (v^1_1)^2=(v^2_1)^2=(v^1_2)^2=(v^2_2)^2=1\hsp {\mbox and}
  \hsp \left[v^\a_k\,,\ps^\b_l\,\right]=0.
$$
In order to define these commutation relations we use
 quantum $(q,s)$-antisymmetrizer (3.15).
 Then, at $q=\pm 1, s=\pm 1$, $\tp^\a_i = v^\a_i\,\ps^\a_i$ are the
components of usual Grassmann bispinor $\tp$.  Making use of the
substitution $\ps^\a_i \rightarrow  \tp^\a_i = v^\a_i\,\ps^\a_i$ into
integral (4.15), one gets
$$
v^2_2\,v^1_2\,v^2_1\,v^1_1\int \bd\tp^1_1 \, \bd\tp^1_2\,\bd\tp^2_1\,
\bd\tp^2_2\, \exp\left\{{\frac 12}
\sum_{\a,\,\b,\,i,\,k}\,b^{\a\b}_{i\,k}\,\tp^\a_i\,\tp^\b_k\right \} =
Pf_{(q,s)}(\hat a).
$$
where  $b^{\a\b}_{i\,k}=a^{\a\b}_{i\,k}\,v^\a_i\,v^\b_k$ (here
summation is absent).

Calculating this integral, we obtain the connection between the usual
Pfaffian of the matrix $\hat b$ and $(q,s)$-Pfaffian (4.14)
$$
v^2_2\,v^1_2\,v^2_1\,v^1_1\,Pf(\hat b)=Pf_{(q,s)}(\hat a).
$$
This relation yields
$$
(Pf_{(q,s)}(\hat a))^2=(Pf(\hat b))^2= det(\hat b).
$$
Hence, we obtain (4.16).

The  $(l,q)$-fermionization of
 the two-dimensional Ising model might be done using
 the lattice $(l,q,s)$-Grassmann field at $q=l=-1, s=1$ (see the
 following section).
In what follows, we just consider  this case, although to compare
 with the case $l=q=1$, one sometimes has not to fix  the value of $l,q$.

Note that for $q=l=-1, s=1$ the $(l,q)$-Grassmann field $\pai{r}$  is
real.  To show this, we define the real field as ${\bar
\k}^{\a}_i(r)=\pai{r}$, where the bar denotes the Hermitian
conjugation. Thus, we have
$$
\overline {(\pai{r}\,\pbj{\rp})} ={\bar \k}^{\b}_j{(\rp)}\, {\bar
\k}^{\a}_i{(r)}= \pbj{\rp}\,\pai{r}\,.
$$
On the other hand, the Hermitian conjugation  of  fifth relation in
(3.34) gives the relation
$$ \k^{2}_i{(r)}\,\k^{1}_j{(\rp)}=-{\frac
1{l^{\ast}}}\,\k^{1}_j{(\rp)}\,\k^{2}_i{(r)}\,.
$$
The consistency condition for these relations leads to the restriction
${\vert l\vert}^2=1$. Using similarly arguments for (3.15), we get
${\vert q\vert}^2={\vert s\vert}^2=1$.

In further calculations we simplify the notation of the lattice
$(l,q)$-Grassmann bispinor field
$$
\ps(r)=
(\,\k^{1}_{1}(r),\,
\k^{2}_{1}(r),\, \k^{1}_{2}(r),\, \k^{2}_{2}(r))\equiv
(\,\ps^1_r,\,\ps^2_r,\,\ps^3_r,\,\ps^4_r\,)\,,
$$
and, as a corollary of (3.15) and (3.34) at  $l=q=-1, s=1$,
its components satisfy
 the commutation relations
 \be
 \lbrack\ps^{1}_{r}\,,\,\ps^{2}_{r}\rbrack=
 \lbrack\ps^{3}_{r}\,,\,\ps^{4}_{r}\rbrack=
 \lbrace\ps^{1}_{r}\,,\,\ps^{4}_{r}\rbrace =
 \lbrace\ps^{2}_{r}\,,\,\ps^{3}_{r}\rbrace=
 \lbrack\ps^{1}_{r}\,,\,\ps^{3}_{r}\rbrack=
 \lbrack\ps^{2}_{r}\,,\,\ps^{4}_{r}\rbrack=0,\tag{4.17}
 \ee
 $$
  \lbrack\ps^{1}_{r}\,,\,\ps^{2}_{r^{\,\prime}}\rbrack=
 \lbrack\ps^{3}_{r}\,,\,\ps^{4}_{r^{\,\prime}}\rbrack=
 \lbrack\ps^{1}_{r}\,,\,\ps^{4}_{r^{\,\prime}}\rbrack =
 \lbrack\ps^{2}_{r}\,,\,\ps^{3}_{r^{\,\prime}}\rbrack=0,
  $$
 \be
   \lbrace\ps^{1}_{r}\,,\,\ps^{3}_{r^{\,\prime}}\rbrace=
 \lbrace\ps^{2}_{r}\,,\,\ps^{4}_{r^{\,\prime}}\rbrace=
 \lbrace\ps^{\a}_{r}\,,\,\ps^{\a}_{r^{\,\prime}}\rbrace=0\,,
 \tag{4.18}
 \ee
 where $\a=1,2,3,4$.

 Using  commutation relations (4.17), (4.18) and
  (4.12)-(4.15),
 it is easy to calculate the one-site integral,
which we need in the following section
  $$
  \int
 \bd\ps_{r}\exp\left\{\sum_{i < k }
 \,b_{ik}\,\ps^{i}_{r_i}\,\ps^{k}_{r_k}+
 b_4\,\ps^{1}_{r_1}\,\ps^{2}_{r_2}\,\ps^{3}_{r_3}\,\ps^{4}_{r_4}\right\}
 \prod ^{4}_{i=1}\,exp({\ps^{i}_{r_i}})
 =
 $$
 \be
 =b_4+Pf_{(l,q)}(b_{ik})+b_{12}+b_{34}+b_{24}+b_{13}-b_{14}-b_{23}
 ,\tag{4.19}
 \ee
 where $i\,,\,k = 1,2,3,4$, $l=q=-1$ and
 $$
 \int \bd\ps_{r}=\int \bd\ps_{1\,r_1}\, \bd\ps_{2\,r_2}\,
 \bd\ps_{3\,r_3}\, \bd\ps_{4\,r_4}=0\,,
 $$
 \be
 Pf_{(l,q)}(b_{ik})=b_{12}\,b_{34}+b_{13}\,b_{24}\,-b_{14}\,b_{23}.
  \tag{4.20}
 \ee

\setcounter{equation}{0}
\section{$(l,q)$-Fermionization of the two-dimensional \hfill\break Ising
model }

In this section we discuss how to express partition
function of the two-dimensional Ising model through
functional integral over the lattice
 $(l,q)$-fermion field  $(l=q=-1)$. The method allows one to  derive
the Lagrangian of the lattice $(l,q)$-fermion field theory describing
the behaviour of the Ising model nearby the critical point.
Our method works for the two-dimensional Ising model of general
type with the only restriction of $Z_2$-invariance of the
 plaquet statistical weight.
 In this case,  the Hamiltonian $h(\sr)$
 (energy per plaquet) includes not only the pairwise
nearest-neighbour interactions, but also
four-spin interaction
\be
h(\sr)=-\sum_{i<j}J_{ij}\,\sr^{i}\,\sr^{j}-J\prod^{4}_{i=1}\sr^{i}
\,,\hsp  i,j=1,\,..\,,4,\tag{5.1}
\ee
where $\sr^{i}=\pm 1$ is the Ising spin,
$ R=(x,y),\hsp x=1,\,..\,,n,\hsp y=1,\,..\,,m,$
denote the plaquet position on the lattice
(as before we use index $r$ for  numbering lattice
sites),
$J_{ij}$ are the parameters of the pairwise interactions and
$J$ is the four-spin coupling constant \hfill\break (see Fig.)

Note that
for the Ising model with only the nearest-neighbours
interactions  we have
$$
J_{1}=J_{34}\,,\hsp
J_{2}=J_{24}\,,\hsp J=J_{12}=J_{13}= J_{24}=J_{14}=0,
$$
where $J_{1}$ and $J_{2}$ are the vertical and horizontal coupling constants,
respectively.

The Ising model Hamiltonian on the whole lattice has the
following form
\be
H=\sum_{\ttr}h(\sr)=-\sum_{\ttr}(\sum_{i<j}J_{ij}\sr^i\,\sr^j+
J\prod^{4}_{i=1}\sr^i)\,.
\tag{5.2}
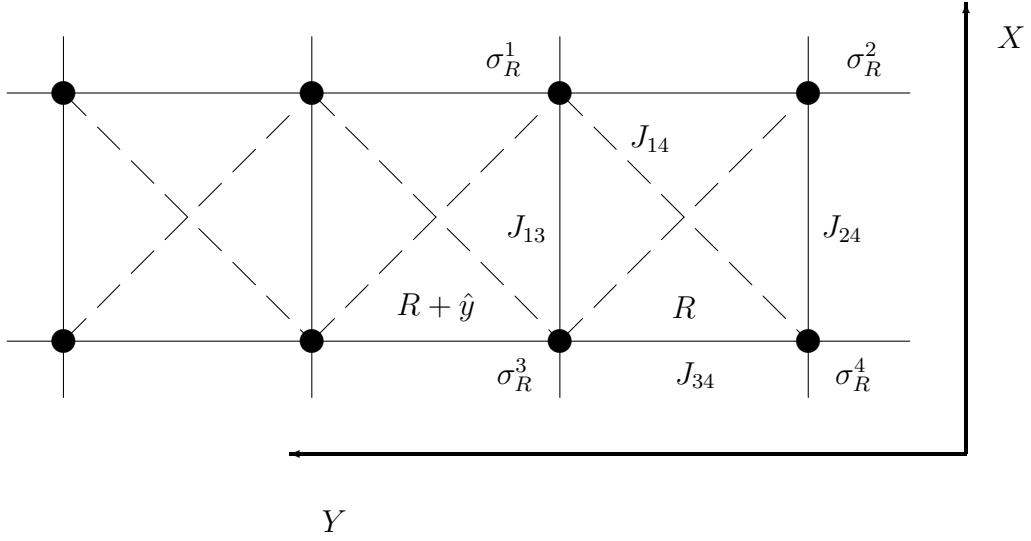
\ee
\begin{figure}[htbp]
\unitlength=1.50mm
\special{em:linewidth 0.4pt}
\linethickness{1.0pt}
\centering
\begin{picture}(100.00,50.00)(60,10)
\put(150.00,10.00){\vector(-1,0){60.00}}
\put(150.00,10.00){\vector(0,1){40.00}}
\put(154.00,47.00){\makebox(0,0)[cc]{$X$}}
\put(94.00,4.00){\makebox(0,0)[cc]{$Y$}}
\emline{145.00}{20.00}{1}{65.00}{20.00}{2}
\emline{70.00}{15.00}{3}{70.00}{47.00}{4}
\emline{92.00}{15.00}{5}{92.00}{47.00}{6}
\emline{114.00}{15.00}{7}{114.00}{47.00}{8}
\emline{136.00}{15.00}{9}{136.00}{47.00}{10}
\emline{145.00}{42.00}{11}{65.00}{42.00}{12}
\put(70.00,42.00){\circle*{2.00}}
\put(92.00,42.00){\circle*{2.00}}
\put(114.00,42.00){\circle*{2.00}}
\put(136.00,42.00){\circle*{2.00}}
\put(70.00,20.00){\circle*{2.00}}
\put(92.00,20.00){\circle*{2.00}}
\put(114.00,20.00){\circle*{2.00}}
\put(136.00,20.00){\circle*{2.00}}
\emline{70.00}{42.00}{13}{72.00}{40.00}{14}
\emline{73.00}{39.00}{15}{75.00}{37.00}{16}
\emline{76.00}{36.00}{17}{78.00}{34.00}{18}
\emline{79.00}{33.00}{19}{81.00}{31.00}{20}
\emline{82.00}{30.00}{21}{84.00}{28.00}{22}
\emline{85.00}{27.00}{23}{87.00}{25.00}{24}
\emline{88.00}{24.00}{25}{90.00}{22.00}{26}
\emline{91.00}{21.00}{27}{92.00}{20.00}{28}
\emline{70.00}{20.00}{29}{72.00}{22.00}{30}
\emline{73.00}{23.00}{31}{75.00}{25.00}{32}
\emline{76.00}{26.00}{33}{78.00}{28.00}{34}
\emline{79.00}{29.00}{35}{81.00}{31.00}{36}
\emline{82.00}{32.00}{37}{84.00}{34.00}{38}
\emline{85.00}{35.00}{39}{87.00}{37.00}{40}
\emline{88.00}{38.00}{41}{90.00}{40.00}{42}
\emline{92.00}{42.00}{43}{94.00}{40.00}{44}
\emline{95.00}{39.00}{45}{97.00}{37.00}{46}
\emline{98.00}{36.00}{47}{100.00}{34.00}{48}
\emline{101.00}{33.00}{49}{103.00}{31.00}{50}
\emline{104.00}{30.00}{51}{106.00}{28.00}{52}
\emline{107.00}{27.00}{53}{109.00}{25.00}{54}
\emline{110.00}{24.00}{55}{112.00}{22.00}{56}
\emline{113.00}{21.00}{57}{114.00}{20.00}{58}
\emline{92.00}{20.00}{59}{94.00}{22.00}{60}
\emline{95.00}{23.00}{61}{97.00}{25.00}{62}
\emline{98.00}{26.00}{63}{100.00}{28.00}{64}
\emline{101.00}{29.00}{65}{103.00}{31.00}{66}
\emline{104.00}{32.00}{67}{106.00}{34.00}{68}
\emline{107.00}{35.00}{69}{109.00}{37.00}{70}
\emline{110.00}{38.00}{71}{112.00}{40.00}{72}
\emline{114.00}{42.00}{73}{116.00}{40.00}{74}
\emline{117.00}{39.00}{75}{119.00}{37.00}{76}
\emline{120.00}{36.00}{77}{122.00}{34.00}{78}
\emline{123.00}{33.00}{79}{125.00}{31.00}{80}
\emline{126.00}{30.00}{81}{128.00}{28.00}{82}
\emline{129.00}{27.00}{83}{131.00}{25.00}{84}
\emline{132.00}{24.00}{85}{134.00}{22.00}{86}
\emline{135.00}{21.00}{87}{136.00}{20.00}{88}
\emline{114.00}{20.00}{89}{116.00}{22.00}{90}
\emline{117.00}{23.00}{91}{119.00}{25.00}{92}
\emline{120.00}{26.00}{93}{122.00}{28.00}{94}
\emline{123.00}{29.00}{95}{125.00}{31.00}{96}
\emline{126.00}{32.00}{97}{128.00}{34.00}{98}
\emline{129.00}{35.00}{99}{131.00}{37.00}{100}
\emline{132.00}{38.00}{101}{134.00}{40.00}{102}
\put(140.00,17.00){\makebox(0,0)[cc]{$\sigma_R^4$}}
\put(126.00,17.00){\makebox(0,0)[cc]{$J_{34}$}}
\put(110.00,17.00){\makebox(0,0)[cc]{$\sigma^3_R$}}
\put(139.00,30.00){\makebox(0,0)[cc]{$J_{24}$}}
\put(141.00,45.00){\makebox(0,0)[cc]{$\sigma^2_R$}}
\put(109.00,45.00){\makebox(0,0)[cc]{$\sigma^1_R$}}
\put(103.00,23.00){\makebox(0,0)[cc]{$R+\hat{y}$}}
\put(125.00,23.00){\makebox(0,0)[cc]{$R$}}
\put(111.00,30.00){\makebox(0,0)[cc]{$J_{13}$}}
\put(122.00,38.00){\makebox(0,0)[cc]{$J_{14}$}}
\end{picture}
\end{figure}

\noindent Here $\hat x$  and $\hat y$  denote the unit vectors along
the horizontal $X$ and vertical  $Y$ axes, respectively.
This model corresponds to the eight-vertex model in external
field
[28],  whose path integral obtained at
the end of the section is not calculable due to the
four-fermion interaction.

Partition function for the model with Hamiltonian (5.2)
can be represented in the form of the product of plaquet statistical
weights
\be
Z=\sum_{\lbrace \si\rbrace}\,e^{-\b H}=\sum_{\lbrace \si\rbrace}
\prod_{\ttr}\,\wrr(\srr\,,\sy\,\vert\,\sx\,,\sxy)\,,\tag{5.3}
\ee
where
$\srr$ is the Ising spin on site $r$
and one gets at $R=r$
\be
\wrr(\srr\,,\sy\,\vert\,\sx\,,\sxy)=
\wrr(\sg 1\,,\sg 2\,\vert\,\sg 3\,,\sg 4)= e^{-\b h}. \tag{5.4}
\ee

It is easy to show that statistical weight (5.4)
can be presented in the following form
\be
\wrr
(\si)=\r(1+\sum_{i\,,\,k}\a_{ik}\sr^i\,\sr^k+\a_4\,\prod^{4}_{i=1}\,\sr^i),
\tag{5.5}
\ee
where
\be
\a_{ik}=<\sr^i\,\sr^k\,>_{h}\,,\hsp
\a_4=<\sg
 1\,\sg 2\,\sg 3\,\sg 4>_{h}\,,\hsp\r=2^{-4}\sum_{\si^1,.,\,\si^4}
 \wrr(\si)\,, \tag{5.6}
\ee
  are expressed through the initial
coupling constants $J_{ik}\,,J$ and
 $<\,f(\si)\,>_{h}$  denotes  averaging  over the
 plaquet statistical weight
$$
 <\,f(\si)\,>_{h}=\sum_{\si^1,.,\,\si^4}f(\si)\wrr(\si)/\sum_{\si^1,.,\,\si^4}
 \wrr (\si).
$$
Note that, choosing the statistical weight in the form
$$
\wrr(\si)=\exp(\K_1\,\sg 3\,\sg 4+\K_2\,\sg 2\,\sg 4
+\K\prod^{4}_{i=1}\sr^i)\,,
$$
where $\K_1=\b J_1\,,\,\,\K_2=\b
J_2\,,\,\, \K=\b J$, one gets the eight-vertex model in zero external
field, and, for the Ising model with only the nearest-neighbours
interactions, one obtains
\be
\wrr(\si)=\exp(\K_1\,\sg 3\,\sg 4+\K_2\,\sg 2\,\sg 4 ).\tag{5.7}
\ee
In the last case, using (5.6), it is easy to calculate the
coefficients in (5.5)
\be
\aligned
\r=\cosh \K_1\cdot \cosh
\K_2\,,\hsp\a_{34}=t_2\,,\hsp\a_{24}=t_1\,,\\
\a_{23}=t_1\,t_2\,,\hsp\a_{14}=\a_{12}=\a_4=0,\endaligned
\tag{5.8}
\ee
where $t_1=\tanh\K_1$  and $t_2=\tanh\K_2$.

To obtain partition function (5.3) as a
functional integral over the lattice $(l,q)$-fermion field, the key
points are: commutation relations
(4.17) and (4.18),  definition of integration rule (4.13) over
$(l,q,s)$-Grassmann field  at $ l=q=-1, s=1$ and representation
 (5.5) for the plaquet statistical weight.
The latter one can be written using (4.19), (4.20)
  in the form of integral
 \begin{eqnarray}
  \wrr(\sg 1\,,\sg 2 \,\vert\, \sg 3\,,\sg 4)&=\r\,\int
 \bd\ps_{\ttr}\exp\left[\,\sum_{i <k}
\,a_{ik}\,\sr^i\,\sr^k\,\ps^{i}_{\ttr}\,\ps^{k}_{\ttr}+\right.
\,\nonumber\\
&\left.
  a_4\,\sr^1\,\sr^2\,\sr^3\,\sr^4\,\ps^{1}_{\ttr}\,\ps^{2}_{\ttr}\,
  \ps^{3}_{\ttr}\,\ps^{4}_{\ttr}\,\right]\,
  \prod^{4}_{i=1}\,\exp\left[\,\ps^{i}_{\ttr}\,\right]\,,
\tag{5.9}
\end{eqnarray}
where
 $ \bd\ps_{\ttr}= \bd\ps_{1\,\ttr}\,
 \bd\ps_{2\,\ttr}\, \bd\ps_{3\,\ttr}\, \bd\ps_{4\,\ttr}\,.  $

The parameters in (5.5) and (5.9) are expressed through each
other as
$$
\aligned
\a_{12}=a_{12},\,\,\a_{13}=a_{13},\,\,\a_{24}=a_{24},\,\,\a_{34}=a_{34},\,\,
\a_{24}=a_{24},\\ \a_{14}=-a_{14},\,\,\a_{23}=-a_{23},\,\,
\a_4=a_4+Pf_{-1}(a_{ik}),\endaligned
$$
and $(l,q)$-fermion field components $\ps$  satisfy commutation
relations (\ref{4.17}) and (\ref{4.18}).

In a similar way, one can write the representation of the
statistical weight product in partition function (5.3)
\be
\prod_{R}\,\wrr(\srr\,,\sy\,\vert\,\sx\,,\sxy)=\r^{nm}\int\,\cD\psi
\exp\left[\,\sum_r\,
\cl_{r}\,(\si\psi)\right]\,\prod_r\,\CP_r\,,\tag{5.10} \ee
where
$$
\cl_{\,r}\,(\si\psi)=\srr\,\sy(a_{12}\,\p1\,\py2+a_{34}\,\p3\,\py4)
+\srr\,\sx(a_{13}\,\p1\,\px3+a_{24}\,\p2\,\px4)+
$$
$$
+\srr\,\sxy\,a_{14}\, \p1\,\pxy4+\sx\,\sy\,a_{23}\,\py2\,\px3+
$$
\be
+a_4(\srr\,\sx\,\sy\,\sxy) (\p1\,\py2\,\px3\,\pxy4),  \tag{5.11}
\ee
$$
 \CP_r=
e^{\p1}\,e^{\py2}\,e^{\px3}\,e^{\pxy4}\,,\hsp
\cD\psi=\prod_{r}\,( \bd\psi_r).
$$
Since not all of the multipliers in
 $\CP_r$ and  $\prod_r\CP_r$  are commutative,
 it is necessary to indicate  their specific ordering.  In
(5.10) we imply the following ordering
$$
\prod_{r}\,\CP_r=\prod^{n}_{y=1}(\prod^{m}_{x=1}\,
\CP_{x,y})=(\CP_{1,1}\cdot
\CP_{2,1}\cdot\cdot\cdot\CP_{m,1}) \times
$$
\be
(\CP_{1,2}\cdot\CP_{2,2}\cdot\cdot\cdot\CP_{m,2})
\cdot\cdot\cdot(\CP_{1,n}\cdot \CP_{2,n}\cdot
\cdot\cdot \CP_{m,n}). \tag{5.12}
\ee
Relation  (5.10)  is proved
by termwise integration over the field components
$\,\p1\,,\,\py2\,,$ $ \px3\,,\,\pxy4\,$.

Product (5.12) can be re-odered in the following way
$$
\prod_{r}\,\CP_r\,\,\Longrightarrow\,\,\prod_r\,Q_r\,,
$$
where
 $$
 Q_r\,=e^{\p4}\,e^{\p3}\, e^{\p2}\, e^{\p1}\,.
 $$
Let us emphasize that, in the course of this re-odering,  we do not
rearrange the non-commuting exponentials, for example, $\exp(\p4)$ and
$\exp(\ppp)$ or $\exp(\p3)$ and $\exp(\pp)$. One can easily check it,
writing down $\CP_r$ in (5.12) explicitly through  the corresponding
exponentials
in (5.11). Right in order to do this re-odering, we need
exotic commutation relations (4.17), (4.18) for the lattice
$(l,q)$-fermion field at $l=q=-1$.  Using this commutation relations and
the nilpotence of the components $\ps^{i}_r$, one transforms $Q_r$ to the
form
$$ Q_r\,=(1+\sum^{4}_{i=1}\,\ps^{i}_r)\exp \lbrack
\cl^{(0)}_{r}\,(\ps)\rbrack,
$$
where $$
\cl^{(0)}_{r}\,(\ps)=\p4\,\p3+\p4\,\p2+\p4\,\p1+\p3\,\p2+\p3\,\p1+\p2\,\p1\,.
$$
As a result, one gets for integral (5.10)
 \be
\prod_{R}\,\wrr =\r^{nm}\int\,\cD\psi
\exp\left[\, \sum_r\,\cl_{r}\,(\si\psi)\right]\,
\prod_r\,\left\{ (1+\sum^{4}_{i=1}\, \ps^{i}_r)\exp \left[
\cl^{(0)}_{r}\,(\ps)\right] \right\} . \tag{5.13}
\ee

The next obvious step is to replace the variables
$$ \psi_{r}^i=\srr\,\fr^i\,,
$$
which leads to vanishing the terms
$\srr\,\si_{r^{\prime}}$  in integral
(5.10). Therefore, the dependence of the Ising spin is localized
in the multiplier
$(1+\sr\sum^{4}_{i=1}\,\fr^{i})$\
$$
1+\sum^{4}_{i=1}\,\psi^{i}_r=1+\srr\,\sum^{4}_{i=1}\,\fr^{i}\,,
$$
$$
\cl_{\,r}\,(\si\psi)=\cl_{\,r}\,(\vr), \hsp \cl^{(0)}_{\,r}\,(\psi)=
\cl^{(0)}_{\,r}\,(\vr)\,.
$$
Now the summation over $\srr$ in partition function (5.3) is
independently done at each lattice site
$$
\sum_{\srr=\pm 1} (1+\srr\,\sum^{4}_{i=1} \,\fr^{i}\,)=2,
$$
and the "bad" multiplier in product (5.13) vanishes.
Since terms in $\cl^{(0)}_{\,r}\,(\vr)$ and  $\cl_{\,r}\,(\vr)$
 are quadratic over the fields $\fr$ (with an additional
 quartic term provided $a_4\neq 0 $) and, therefore, commutative, one
obtains
 \be
 Z =(2\r)^{nm}\int\,\cD\vr \exp\left[\,
\sum_r(\cl^{(0)}_{\,r}\,(\vr) +\cl_{\,r}\,(\vr))\right]\,,
\tag{5.14}
\ee
where
$$
\cl^{(0)}_{\,r}\,(\vr)=\f4\,\f3+\f4\,\f2+\f4\,\f1+\f3\,\f2+\f3\,\f1+
\f2\,\f1\,,
$$
$$
\cl_{\,r}\,(\vr)=a_{12}\,\f1\,\fy2+a_{34}\,\f3\,\fy4+
a_{13}\,\f1\,\fx3+ a_{24}\,\f2\,\fx4+
$$
$$
+a_{14}\,\f1\,\fxy4+
a_{23}\,\f2\,\fxy3+ a_4\,\f1\,\fy2\,\fy3\,\fxy4\,.
$$
At $a_4=0$,  one can represent functional integral (5.14) as
Gaussian integral
\be
Z =(2\r)^{nm}\int\,\cD\vr \exp\left[\, {\frac 12}
\sum \f{i}\,D^{i\,j}_{r,\,r^{\,\prime}}\,\rf{j}
\right]\, ,\tag{5.15}
\ee
where
$$
\hat D=
\bacc
0 &
1+a_{12}\,\nx &  a_{13}\,-\mx  & a_{14}\,\ny+\mx\\
\vspace{3truemm}
1+a_{12}\,\mx &  0 & a_{23}\,\mmy+ \mx  & a_{24} -\nx\\
\vspace{3truemm}
a_{13}\,+\nx  & a_{23}\,\ny+\nx  & 0 & 1+a_{34}\,\ny\\
\vspace{3truemm}
a_{14}\,\mmy+\nx & a_{24}\,+\mx & 1+a_{34}\,\mmy  & 0\eacc\,.
$$
This integral is expresed through $(l,q)$-Pfaffian of the matrix
$\hat D$.
 Similarly to the case of usual
fermionization of the two-dimensional Ising model (1.2)-(1.4), this
matrix is linear in the one-step shift operators $ \nx\,,\,\ny$.
In fact the accounting of the boundary condition in (5.15) gives
the $(l,q)$-deformed counterparts $(l=q=-1)$ of (1.2), (1.3).
For the sake of simplicity, we consider only $Z^{AA}$, where index
"A"  again denotes the antiperiodic boundary conditions along the axes
$X$ and $Y$. When turning to the momentum representation, the momentum
components $p_1, p_2$ take half-integer values.

In what follows, one restricts himself to considering the
 two-dimen\-sional Ising model with only the nearest-neighbours
interactions. In this case, the coefficients
 $a_{ij}$ are written in (5.8) and functional integral (5.15)
in momentum representation can be rewritten in the following
form
\be
Z=(2\r)^{nm}\int\cD\vr\exp\lbrace\frac 12
\sum_{p_1,p_2}\vr^i \mm\, D^{i\,j} \qq\,\vr^j \qq \rbrace\,,
 \tag{5.16}
\ee
 where
$$
D^{i\,j}\qq=\bacc
0&1&-\mp&\mp\\ \vspace{3truemm}
1&0& \ip - t_1t_2\mq & t_1-\ip\\ \vspace{3truemm}
\ip& \mp -t_1t_2\iq &0&1+t_2\iq \\ \vspace{3truemm}
\ip&t_1+\mp&1+t_2\mq&0 \eacc\,,
$$
and
\begin{eqnarray}
\cD\vr=\prod_{p_i>0}&\bd Re\vr_1 \qq\, \bd Re\vr_2 \qq\,
\bd Re\vr_3 \qq \, \bd Re\vr_4 \qq\,\nonumber\\
&\bd Im\vr_1 \qq\, \bd Im\vr_2 \qq\,  \bd Im\vr_3 \qq\,  \bd Im\vr_4
\qq\nonumber\,.
 \end{eqnarray}
 At $q=-1, s=1$, using (4.16) for the calculation
of the $(q,s)$-deformed Gaussian functional integral, one immediately
gets integral (5.16). As a result, one obtains known expression (1.6) for
$Z^{AA}$.

Thus, (5.16) connects the partition function of
 the two-dimensional Ising model with functional
integral in the lattice free $(l,q)$-fermion field theory $(l=q=-1)$.
The coincidence of the results of calculations (5.16) and (1.8)
allows one to establish the validity of the definiton
of the Berezin integral for the lattice $(l,q,s)$-Grassmann
field, proposed in (4.13).

Note that in paper [29]  it was demonstrated the connection between
paragrassmann algebras and the some representation  spaces of quantum
matrix group $L_q(2,C)$ with  $q$ being a $(p+1)$-root of unity where
$p$ is integer ($\theta ^{p+1}=0$ for paragrasmann $\theta$).
In our case, using definition of the grassmann anyonic field given in [30 ],
the commutation relations (4.17)-(4.18)
 for the lattice  $(l,q,s)$-fermion field  $(l=q=-1,s=1)$
can be considered as definition of the lattice
two-component paragrassmann anyonic field  with $l=e^{i\pi\nu}, (\nu=1)$
and $p=3$.

 We would like to thank  A.~Mironov for reading manuscript and
critical remarks. Also we acknowledge V.~Akulov, A.~Mikhailov,
 A.~Losev, D.V.Volkov, A.~Zabrodin for critical remarks. We
are very grateful to V.~Rubtsov for the help during the work.

V.S. thanks
Prof. A.Niemi for the hospitality and the exellent conditions at the
University of Uppsala, where this paper has been finished.

We would
like to thank the members of Bogoluybov Institute for Theoretical
Physics and  Institute of Mathematics of the  Ukrainian Academy of
Science for the stimulating discussions.

The work is
partially supported by grant 93-02-14365 of the Russian
Foundation of Fundamental Research and by grant MGK000 of the International
Soros Foundation.

\end{document}